\documentclass[aps,prd,twocolumn,groupedaddress,nofootinbib]{revtex4-1}
\usepackage{tensor, amsmath, amssymb, graphics, geometry, mathtools, graphicx, wrapfig}
\usepackage[
	colorlinks=true,
	linkcolor=blue,
	citecolor=blue,
	filecolor=magenta,
	urlcolor=blue
]{hyperref}
\newcommand{\barT}{{\overline{T}}}

\begin{document}

\title{Exploration of Parameters in \textit{f}(\textit{R},\textit{T}) Gravity and Comparison with Type Ia Supernovae Data}
\author{Vincent R. Siggia}
\author{Eric D. Carlson}
\email{ecarlson@wfu.edu}
\author{P. Lee Pryor}
\affiliation{Department of Physics, Wake Forest University, 1834 Wake Forest Road, Winston-Salem, North Carolina 27109, USA}

\begin{abstract}
    We consider Type Ia supernova data in the context of $f(R,T)$ gravity. We consider functions of the form $f(R,T)=R+\lambda T^\epsilon$ where $\epsilon<1$. We find that for all models with $\epsilon<0$, the Universe transitions to exponential growth at late times, just as it does in the standard cosmological model, which corresponds to $\epsilon=0$.  It also fits the type Ia supernova data slightly better than the standard cosmological model, without increasing the number of parameters of the theory. In contrast, the fits for $\epsilon >0$ rapidly become worse than the standard cosmological model.
\end{abstract}

\maketitle

\section{Introduction}
It is generally acknowledged that the Universe's expansion is accelerating, as first observed using Type Ia supernovae (SNe Ia) \cite{Riess_1998,Perlmutter_1998,Perlmutter_1999}, and later demonstrated using the cosmic microwave background radiation \cite{Planck2018} and baryon acoustic oscillations \cite{Alam_2017}. The conventional explanation is that this is due to a cosmological constant, $\Lambda$. This, with the indication that most of the nonrelativistic matter in the Universe is cold dark matter (CDM) and the added assumption that the universe is spatially flat, leads to the standard $\Lambda$CDM model of the Universe.

Although the $\Lambda$CDM model fits the data well, there has been enormous interest in alternative theories, including modifications of gravity. For example, in $f(R)$ gravity the curvature term, $R$, is replaced by some function of the curvature scalar \cite{10.1093/mnras/150.1.1}. These models have been used to try to explain observational properties of the current universe, such as SNe Ia data, the CMB and baryon acoustic oscillations (BAO) \cite{GameOver,odintsov2025dynamicaldarkenergyfr}.  Harko \textit{et al.} proposed an alternative where $R$ is replaced by $f(R,T)$, a function that also depends on the trace of the stress-energy tensor $T$ \cite{Harko_2011}. 

In a previous paper \cite{Siggia_2025}, we focused on a modification of the standard Lagrangian by replacing the scalar curvature, $R$, by some combination of a function of the curvature and the stress energy tensor such that $R \rightarrow f(R,T)$. We specifically focused on the case where $f$ is additively separable, that is, $f(R,T) = f_1(R) + f_2(T)$, and specialized to
\begin{equation}\label{f(R,T)}
    f(R,T) = R + \lambda T^\epsilon\;.
\end{equation}
Thus, our action is
\begin{equation}\label{Action}
    S = \int d^4x \sqrt{-g}\left[\mathcal{L}_m -\frac{1}{2\kappa^2}\big(R+\lambda T^\epsilon\big) \right] \; ,
\end{equation}
where $\mathcal{L}_m$ is the Lagrangian for ordinary and dark matter, and $\kappa^2=8\pi G$. The previous paper focused on the value $\epsilon=-1$; in this paper we will explore the full range of $\epsilon<1$.

The goal of the $\lambda T^\epsilon$ term is to create a universe that behaves in a typical matter- or energy-dominated manner until late times, at which point the new term takes over, possibly explaining the accelerated expansion. To ensure a universe that changes its behavior only at late times, the new term must dominate $\mathcal{L}_m$ at low energy density. This criterion is satisfied if $\epsilon < 1$. Note that $\epsilon = 0$ corresponds to the standard $\Lambda$CDM Universe with $\lambda = 2\Lambda$.

The trace of the stress-energy tensor, $T$, is normally computed beginning with the matter Lagrangian, $\mathcal{L}_m$. This is either the standard-model Lagrangian or whatever extension is necessary to include dark matter. Due to the complex or unknown nature of this Lagrangian, it has become standard practice to naïvely assume the stress-energy tensor takes the form appropriate for a perfect fluid. In standard general relativity, the procedure was worked out in \cite{Brown_1993}. However, there are subtleties in applying this formalism in $f(R,T)$ gravity that have been often ignored in many articles \cite{Harko_2011,S_ez_G_mez_2016,PhysRevD.94.084052,PhysRevD.95.123536,Carvalho_2017,Deb_2018,RUDRA2021115428,Feng}, which were later pointed out by
\cite{Minazzoli_2012,NOJIRI20171} and further elaborated by \cite{Fisher_2019}. For further discussion on this issue, see \cite{Fisher_2019}.

In Sec.~\ref{sec:General}, the general framework of our model is outlined, resulting in the modified Friedmann equations. The scale factor of the Universe is derived in asymptotic limits for multiple values of $\epsilon<1$ in Sec.~\ref{sec:Scale}. Then in Sec.~\ref{sec:Data}, the Pantheon dataset \cite{Scolnic_2018,Lu_2022,Abbott_2019} is used to compare our predictions for various $\epsilon<1$ against $\Lambda$CDM. In Sec.~\ref{Sec:Conclusion} we summarize our results.

We will use units where $\hbar = c = 1$, and we have the mostly minus sign convention for the metric. The Riemann tensor is given by ${R^\alpha}_{\mu\beta\nu} = \nabla_\beta \Gamma^\alpha_{\mu\nu} - \cdots$ and the Ricci Tenor by $R_{\mu\nu} = {R^\alpha}_{\mu\alpha\nu}$.

\section{General Model}\label{sec:General}

The Einstein equations for our model are derived by varying the action Eq.~(\ref{Action}) with respect to $g^{\mu\nu}$ yielding
\begin{equation}\label{Einstein EQ}
   R_{\mu \nu }-\frac{1}{2}R\,g_{\mu \nu }=\kappa ^2T_{\mu \nu }+\frac{1}{2}\lambda T^\epsilon g_{\mu \nu }-\epsilon\lambda T^{\epsilon-1
   }\frac{\partial\,T}{\partial g^{\mu \nu }}\;.
\end{equation}
Taking the divergence of Eq.~(\ref{Einstein EQ}) shows that the stress-energy is not conserved \cite{Siggia_2025,PhysRevD.90.028501},
\begin{equation}
   \kappa^2\nabla^\mu T_{\mu\nu}= \nabla^\mu\left(\epsilon\lambda T^{\epsilon-1}\frac{\partial\,T}{\partial g^{\mu \nu }}\right)-\frac{1}{2}\epsilon\lambda T^{\epsilon-1}\nabla_\nu T\;.
\end{equation}

In general, perfect fluids are described in terms of the particle number density $n$ and entropy per particle $s$; therefore, the naïve energy density and pressure in a comoving frame can be written as $\rho=\rho(n,s)$ and $p=p(n,s)$. Dealing with perfect fluids is difficult because of the necessity of using Lagrange multipliers to enforce conservation laws, as shown by \cite{Brown_1993} and further explored in the context of $f(R,T)$ gravity by \cite{Fisher_2019}. In this approach,  the number density $n$ and four velocity $u^\mu$ are combined into a single unconstrained current density $J^\mu=n u^\mu$, where now $n$ is treated as a dependent quantity given by $n=\sqrt{g_{\mu\nu}J^\mu J^\nu}$. The current density $J^\mu$ is not conserved; instead, one finds
\begin{equation}\label{General Current Density}
    \nabla_\mu\left[\left(1+\frac{2\epsilon\lambda }{\kappa ^2}T^{\epsilon-1}\right)J^\mu\right]=0\;,
\end{equation}
where the expression inside the square brackets can be considered an effective current density $J'^\mu$. 

In standard gravity, because the stress-energy tensor is conserved, it can be shown that the energy density and pressure are related by
\begin{equation}
    n\frac{\partial\rho}{\partial n}=\rho+p.
\end{equation}
In general, this does not necessarily hold. However, this can still be considered a definition of the naïve pressure $p$.

The stress-energy quantities appearing in Eq.~(\ref{Einstein EQ}) become\begin{subequations}
    \begin{align}
         \label{General Stress-Energy}T_{\mu \nu }&=(\rho +p)u_{\mu }u_{\nu }-g_{\mu \nu }\mathcal{L}_m \; , \\
        \label{General Stress-Energy Trace}T&=(\rho +p)-4\mathcal{L}_m\;, \\
        \label{T Variation}\frac{\partial T}{\partial g^{\mu\nu }}&=-\frac{1}{2}u_{\mu }u_{\nu }\left(4+n\frac{\partial }{\partial n}\right)(\rho +p)\;.
    \end{align}
\end{subequations}

Unfortunately, the matter Lagrangian still contains Lagrange multipliers. All quantities can be calculated ``on-shell''; chosen to satisfy the equations of motion in order to eliminate the Lagrange multiplier terms. We denote such quantities by an over-bar. Following the analysis of \cite{Siggia_2025}, eliminating these Lagrange multipliers leads to the on-shell matter Lagrangian
\begin{equation}\label{lm-bar}
    \overline{\mathcal{L}_m}=p+\frac{\epsilon\lambda}{2\kappa ^2\barT^{1-\epsilon}+4\epsilon\lambda}n\frac{\partial }{\partial n}(\rho +p)\;.
\end{equation}
Simply replacing $\mathcal{L}_m$ with $\overline{\mathcal{L}_m}$ in Eqs.~(\ref{General Stress-Energy})-(\ref{T Variation}) produces their on-shell equivalents. In particular, by combining Eqs.~(\ref{General Stress-Energy Trace}) and (\ref{lm-bar}), the on-shell stress-energy trace $\barT$ becomes an implicit function of $\rho$ and $p$ as
\begin{equation}\label{General Quad}
    \rho-3p=\barT+\frac{2\lambda\epsilon }{\kappa ^2}\barT^{\epsilon }+\frac{2\lambda\epsilon }{\kappa ^2}\barT^{\epsilon -1}\left(4+n\frac{\partial }{\partial n}\right)p\;.
\end{equation}

Assuming a flat Friedmann-Lemaître-Robinson-Walker (FLRW) metric, 
\begin{equation}\label{FLRW Metric}
    ds^2=d t^2-a^2(t)\left(d r^2+r^2d\Omega^2\right)\;,
\end{equation}
the equations of motion in terms of $n$ and $s$ are
\begin{subequations}
\begin{align}
    \nonumber3\left(\frac{\dot{a}}{a}\right)^2&=\frac{\kappa^2}{4}\left(3\rho+3p+\barT\right)+\frac{\lambda}{2}\barT^\epsilon\\
   &+\frac{\epsilon\lambda}{2}\barT^{-(1-\epsilon)}\left(4+n\frac{\partial}{\partial n}\right)(\rho+p)\;,\label{tt-Friedmann}\\
    \nonumber-6\left(\frac{\ddot{a}}{a}\right)&=\frac{\kappa^2}{2}\left(3\rho+3p-\barT\right)-\lambda\barT^\epsilon\\
    &+\frac{\epsilon\lambda}{2}\barT^{-(1-\epsilon)}\left(4+n\frac{\partial}{\partial n}\right)(\rho+p)\,.\label{rr-Friedmann}
\end{align}
\end{subequations}

\section{The Scale Factor}\label{sec:Scale}

The discovery of SNe Ia data indicating that the expansion of the Universe was currently accelerating led to the reintroduction of the cosmological constant $\Lambda$ \cite{Riess_1998}. However, many alternatives have been considered. As we demonstrated in our previous paper \cite{Siggia_2025}, we can achieve this in the case $\epsilon = -1$. How general is this? We will again assume that the space metric is flat, given by Eq.~(\ref{FLRW Metric}). Due to being derived from the matter Lagrangian $\mathcal{L}_m$, during the nonrelativistic era, we still expect the pressure $p=0$ and the naïve energy density $\rho\propto n$ \cite{Siggia_2025}.

Using $p=0$, Eq.~(\ref{General Quad}) reduces to
\begin{equation}\label{Quadratic}
    \rho=\barT+\frac{2\epsilon\lambda}{\kappa ^2}\barT^\epsilon\;,
\end{equation}
which can be solved for $\barT$. Equations~(\ref{tt-Friedmann}) and (\ref{rr-Friedmann}) in terms of $\barT$ become
\begin{subequations}
\begin{align}  
    \label{1st P=0}&\frac{\dot{a}^2}{a^2}=\frac{\kappa^2\barT}{3}\left[1+\frac{(8\epsilon+1)\lambda}{2\kappa^2\barT^{1-\epsilon}}+5\bigg(\frac{\epsilon\lambda}{\kappa^2\barT^{1-\epsilon}}\bigg)^2\right],\\
    \label{2nd P=0}&\frac{\ddot{a}}{a}=-\frac{\kappa^2\barT}{6}\left[1+\frac{(11\epsilon-2)\lambda}{2\kappa^2\barT^{1-\epsilon}}+5\bigg(\frac{\epsilon\lambda}{\kappa^2\barT^{1-\epsilon}}\bigg)^2\right].
\end{align}
\end{subequations}
Using Eq.~(\ref{Quadratic}), the effective current density reduces to
\begin{equation}
    J'^{\mu }=\left(1+\frac{2\epsilon\lambda }{\kappa ^2}\barT^{\epsilon-1}\right)J^{\mu }=\frac{\rho }{\barT}\left(n \, u^{\mu }\right)\;.
\end{equation}
In the nonrelativistic present Universe, since $n$ and $\rho$ are both derived from the matter Lagrangian $\mathcal{L}_m$, we would have $\rho \propto n$, so that $J'^0 \propto \rho^2/\barT$ will be conserved. This implies that $\rho^2a^3/\barT$ is constant, which we write as
\begin{equation}\label{Constant Conservation}
   \frac{\rho ^2}{\barT}a^3=\frac{12N^3}{\kappa^2}\;,
\end{equation}
where the constant was chosen for later comparison and $N$ is an arbitrary normalization factor \cite{Siggia_2025}. 
With the help of Eq.~(\ref{Quadratic}), Eq.~(\ref{Constant Conservation}) and its derivative yield the relations
\begin{align}
    \label{A}a&=N\Bigg[\frac{12\kappa^2\barT^{1-2\epsilon}}{\big(\kappa^2\barT^{1-\epsilon}+2\epsilon\lambda\big)^2}\Bigg]^{1/3}\; , \\
    \label{DotA/A}\frac{\dot{a}}{a}&=-\frac{\dot{\barT}\big[\kappa^2\barT^{1-\epsilon}-(1-2\epsilon)2\epsilon\lambda\big]}{3\barT\big(\kappa^2\barT^{1-\epsilon}+2\epsilon\lambda\big)}\;.
\end{align}

We rescale time and the stress-energy trace to produce dimensionless quantities $\tau$ and $x$ defined by
\begin{subequations}
\begin{align}
    &\tau =t \sqrt{\frac{\kappa^2}{6} \left(\frac{2\lambda }{\kappa ^2}\right)^{\frac{1}{1-\epsilon}}}\; , \label{t scaling} \\
    &\barT(t)=x(\tau)\left(\frac{2\lambda }{\kappa ^2}\right)^{\frac{1}{1-\epsilon}}\; . \label{T Scaling}
\end{align}
\end{subequations}
Combining Eqs.~(\ref{DotA/A}) and (\ref{1st P=0}) yields the differential equation
\begin{align}\label{diff X}
    \nonumber\frac{dx}{d\tau}=-3&\frac{x^{1-\epsilon}+\epsilon}{x^{1-\epsilon}-(1-2\epsilon)\epsilon}\\
    &\times\sqrt{\frac{4x^{2(1-\epsilon)}+(8\epsilon+1)x^{1-\epsilon}+5\epsilon^2}{2x^{-(1+2\epsilon)}}}\; .
\end{align}

\subsection{Case: $\epsilon<0$}

First, consider the case $\epsilon<0$. During the matter-dominated Universe, the energy density is approximately infinite, i.e. $x\rightarrow\infty$. As the Universe expands, the energy density will decay until there is only the vacuum of spacetime. Therefore, by Eq.~(\ref{diff X}), $x\rightarrow(-\epsilon)^{\frac{1}{1-\epsilon}}$. In these limits, solving Eq.~(\ref{diff X}) determines the early and late behavior of the unitless stress-energy trace $x$ as
\begin{equation}\label{X}
    x=
    \begin{cases} 
      \frac{2}{9\tau^{2}} & \tau\rightarrow 0\;, \\
      (-\epsilon)^{\frac{1}{1-\epsilon}}+\alpha_\epsilon \exp\left[-\frac{3}{2}\sqrt{(-\epsilon)^{\frac{\epsilon}{1-\epsilon}}\left(\frac{1-\epsilon}{2}\right)}\tau\right] & \tau\rightarrow\infty\;, 
    \end{cases}
\end{equation}
where $\alpha_\epsilon$ can be determined numerically for each $\epsilon$. This gives an asymptotic behavior of the scale factor as
\begin{equation}\label{Limit A}
    a(t)=N
    \begin{cases} 
      (3t)^{2/3} & H_\lambda  t\rightarrow 0\;, \\
     H_\lambda^{-2/3} \beta_\epsilon\exp(H_\lambda t) &  H_\lambda t\rightarrow\infty\;,
    \end{cases}
\end{equation}
where $\beta_\epsilon = \left[\alpha_\epsilon^{-2} (1-\epsilon)^{-1}(-\epsilon)^\frac{1+\epsilon}{1-\epsilon} \right]^{1/3}$ and $H_\lambda$ is defined by Eq.~(\ref{1st P=0}) at late times as
\begin{equation}\label{HLambda}
    H_\lambda^2=\frac{1-\epsilon}{6}\left[\left(-\frac{2\epsilon}{\kappa^2}\right)^{\epsilon}\lambda\right]^{\frac{1}{1-\epsilon}}\;.
\end{equation}
In the limit of $\epsilon\rightarrow0$, i.e. $\Lambda$CDM,  Eq.~(\ref{Limit A}) still holds, where $\beta_\epsilon=1$ and Eq.~(\ref{HLambda}) becomes $H_\Lambda^2=\lambda/6=\Lambda/3$.

The Hubble parameter as a function of unitless time $H_\lambda t$ is shown in Fig.~\ref{fig:H_Constant}. However, as in our previous paper, there is no reason to assume that the asymptotic Hubble constants $H_\Lambda$ and $H_\lambda$ should match, since the goal of these models is not to make the Universe with particular future behavior, but instead to match the observed redshift-luminosity curves \cite{Siggia_2025}. 

Assuming the limit of Eq.~(\ref{HLambda}) as $\epsilon\rightarrow-\infty$ makes sense, the differential Eq.~(\ref{diff X}) can be solved exactly. This yields
\begin{equation}\label{Xinf}
    x=
    \begin{cases} 
      \frac{2}{9\tau^{2}} & \tau\leq\frac{\sqrt{2}}{3}\;, \\
      \left\{-\epsilon\left[1+\frac{1}{2}\cosh{\big(\frac{3\tau}{\sqrt8}-\frac{1}{2}\big)}\right]\right\}^\frac{1}{1-\epsilon} & \tau>\frac{\sqrt{2}}{3}\;,
    \end{cases}
\end{equation}
where we have delayed completing the limit $\epsilon \rightarrow - \infty$. Using Eq.~(\ref{1st P=0}), and defining $H_\infty=\lim_{\epsilon \rightarrow -\infty} H_\lambda$, the scale factor is determined to be
\begin{equation}\label{Ainf}
    a=N
    \begin{cases} 
        (3t)^{2/3} & H_\infty t\leq\frac{1}{3}\;, \\
        H_\infty^{-2/3}\left[1+2\sinh\big(\frac{3}{2}H_\infty t-\frac{1}{2}\big)\right]^{2/3}& H_\infty  t>\frac{1}{3}\;.
    \end{cases}
\end{equation}

\subsection{Case: $\epsilon\in\left(0,1\right)$}

For $\epsilon\in\left(0,1\right)$, assuming $\lambda>0$, it can be shown from Eq.~(\ref{2nd P=0}) that the Universe will have no period of acceleration unless $\epsilon<\frac{1}{41}(22-8\sqrt{5})\approx0.1003$. However, there is a problem for $\epsilon\in\left(0,\frac{1}{2}\right]$; namely, there are no real solutions to Eqs.~(\ref{Quadratic}) and (\ref{Constant Conservation}) combined for sufficiently large values of $a$. This suggests that the theory becomes inconsistent once the number density gets sufficiently low. 

\begin{figure}
\includegraphics[width=\columnwidth]{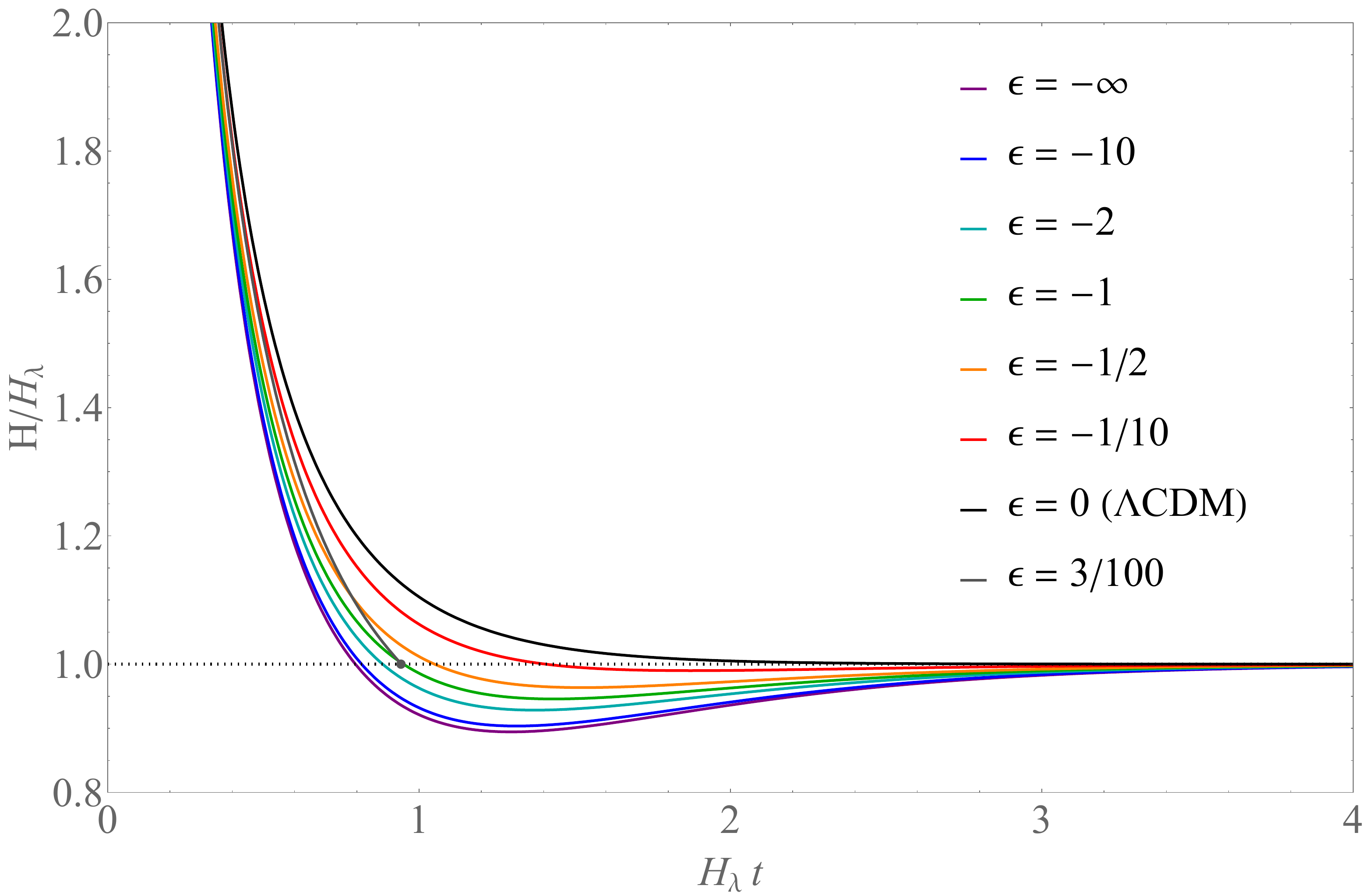}
\caption{\label{fig:H_Constant} The Hubble Parameter as a function of unitless time is shown here.  Note that for $\epsilon=0.03$, the solution breaks down at an endpoint. This behavior occurs when $H_\lambda$ attains the value determined by Eq.~(\ref{Break Down H}) when $\epsilon\in\left(0,\frac12\right]$.}
\end{figure}

We can nonetheless analyze this case, but simply constrain $\tau_0$ such that $\tau_0$ occurs before this critical failure point. For quite small values of $\epsilon$, the best fit to the data occurs when we take $\tau_0$ to be this limiting value. Unfortunately, the fit quickly becomes very poor. The breakdown point occurs where the Hubble parameter takes the value
\begin{equation}\label{Break Down H}
    H_\lambda^2=\frac{1}{12}\left[\frac{2\epsilon^\epsilon\lambda}{\kappa^{2\epsilon}(1-2\epsilon)^{1-2\epsilon}}\right]^\frac{1}{1-\epsilon}(1-\epsilon)(1+16\epsilon-16\epsilon^2)\;.
\end{equation}
This behavior is illustrated in Fig.~\ref{fig:H_Constant} for the case $\epsilon=0.03$.

If $\lambda <0$, one might naïvely expect from Eq.~(\ref{2nd P=0}) that a period of accelerated expansion could occur if $\epsilon>\frac{1}{41}(22+8\sqrt{5})\approx0.9729$. However, comparison with Eq.~(\ref{1st P=0}) shows that the Universe would instead expand to a maximum size followed by an era of collapse without ever attaining accelerating expansion.

In the limit where $\epsilon\rightarrow1$, Eqs.~(\ref{1st P=0}) and (\ref{2nd P=0}) are equivalent to the standard flat Friedmann equations without a cosmological constant; the only difference being a rescaling of the constants.

\section{Luminosity Distance and Observation}\label{sec:Data}

The redshift and luminosity distance can be understood through the relationship
\begin{equation}\label{Initial Luminosity Distance}
    d_L=a_0(1+z) \int_t^{t_0}\frac{d t'}{a(t')}\;,
\end{equation}
where the zero subscript denotes the quantity at present time. Similarly for the $f(R,T)$ theory considered, by combining the Eqs. (\ref{DotA/A})-(\ref{diff X}) and evaluating at the present time, 
\begin{equation}\label{Time/Tau}
    H_0\frac{dt}{d\tau}=\sqrt{\frac{4x_0^{2(1-\epsilon)}+(1+8\epsilon)x_0^{1-\epsilon}+5\epsilon^2}{2x_0^{1-2\epsilon}}}\;,
    \end{equation}
where $x_0=x(\tau_0)$, a similar relation to Eq.~(\ref{Initial Luminosity Distance}) can be derived. By inserting Eqs.~(\ref{A}) in terms of $x(\tau)$ and (\ref{Time/Tau}) into (\ref{Initial Luminosity Distance}) yields
\begin{align}
    \nonumber &H_0 d_L=(1+z)\left[\frac{4x^{2(1-\epsilon)}_0+(1+8\epsilon)x^{1-\epsilon}_0+5\epsilon^2}{2x^{1-2\epsilon}_0}\right]^{1/2}\times\\
     &\;\left[\frac{x^{1-2\epsilon}_0}{(x^{1-\epsilon}_0+\epsilon)^2}\right]^{1/3}\int_\tau^{\tau_0}\left[\frac{(x(\tau')^{1-\epsilon}+\epsilon)^2}{x(\tau')^{1-2\epsilon}}\right]^{1/3}d\tau'\;,
\end{align}
and
\begin{equation}
    1+z=\left[\frac{x_0^{1-2\epsilon}}{(x_0^{1-\epsilon}+\epsilon)^2}\right]^{1/3}\left[\frac{(x^{1-\epsilon}+\epsilon)^2}{x^{1-2\epsilon}}\right]^{1/3}\;.
\end{equation}
The distance modulus $\mu$ can be easily calculated from the luminosity distance $d_L$, using
\begin{equation}
    \mu=5\,\log_{10}\left(\frac{d_L}{10\,\text{pc}}\right)\;.
\end{equation}

\begin{figure}
\includegraphics[width=\columnwidth]{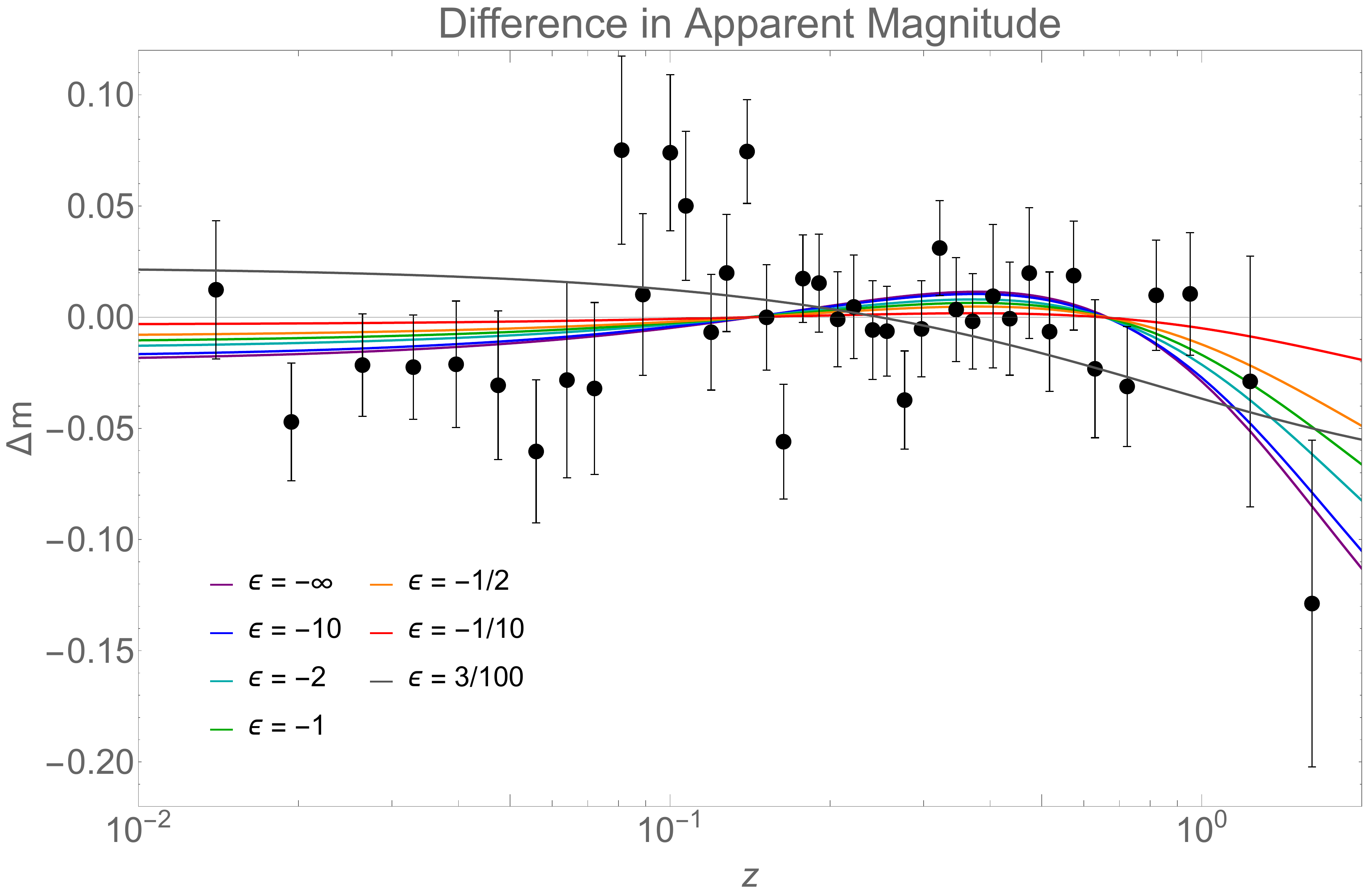}
\caption{\label{fig:Apparent_Magitnude} The difference $\Delta m=m_f-m_\Lambda$ between $f(R,T)$ apparent magnitude and the best fit standard $\Lambda$CDM apparent magnitude. The Pantheon binned data is shown here to reduce visual clutter; provided by \cite{Scolnic_2018}. Both the statistical analysis and fitting were done with the complete dataset.}
\end{figure}

 The Pantheon dataset \cite{Scolnic_2018,Lu_2022,Abbott_2019} contains data from 1048 supernovae. We have included Fig.~\ref{fig:Apparent_Magitnude} to highlight the difference in apparent magnitude between the best fit $\Lambda$CDM and $f(R,T)$ predictions for each model.

In order to find the best fit values for our models, we used standard $\chi^2$ minimization with respect to $M$, given by
\begin{equation}
    \chi^2=\sum_{i=1}^{n}\left(\frac{\mu(z_i)-m_i+M}{\sigma_i}\right)^2\;,
\end{equation}
where $M$, $m_i$, $z_i$, and $\sigma_i$ are the absolute magnitude, the apparent magnitude, the corrected redshift, and the error in the apparent magnitude of the supernova data. We then found a minimum for each model by scanning through various values of $\tau_0$. By combining Eqs.~(\ref{t scaling}) and (\ref{Time/Tau}), the value of $\lambda$ can be written as
\begin{equation}
    \lambda=\frac12\kappa^{2\epsilon} \left(12H_0^2\bar{\lambda}\right)^{1-\epsilon}\;,
\end{equation}
where
\begin{equation}
    \bar{\lambda}=\frac{x_0^{1-2\epsilon}}{4x_0^{2(1-\epsilon)}+(1+8\epsilon)x_0^{1-\epsilon}+5\epsilon^2} \; .
\end{equation}
We note that in the special case $\epsilon = 0$, $\bar \lambda = \Omega_\Lambda$ where $\Omega_\Lambda$ is the dark energy parameter. Our value for $\Omega_\Lambda$ agrees with \cite{Scolnic_2018} since we both are only analyzing SNe Ia data. For the case $\epsilon = 0.03$ the best fit occurs at the critical failure point, so that $\bar \lambda$ is forced to be no larger than the indicated value. We did not include an error in the table because the fit is already so poor that this case does not deserve more consideration.

In Tab.~\ref{tab:DataTable}, $\beta_\epsilon$ and the best fit for $H_\lambda t_0$ and $\bar \lambda$ with errors are shown for seven representative values of $\epsilon$, along with $\chi^2$ values for the best fit. The central value for $M$ is shown using an assumed Hubble parameter $H_0=71.5$ km/s/Mpc. Note that the best fit for $M$ is almost independent of model, but recall that the value is sensitive to the actual value chosen for $H_0$. Similarly to the flat $\Lambda$CDM model, the absolute magnitude $M$ for SNe Ia and Hubble's constant $H_0$ are degenerate parameters \cite{Scolnic_2018}. As in our previous paper \cite{Siggia_2025}, the second parameter used in all theories is $\lambda$. Therefore, there are 1046 degrees of freedom.

\begin{table}
\begin{center}
\begin{tabular}{|c||c|c|c|c|c|} 
 \hline
 $\epsilon$ & $\beta_\epsilon$ & $H_\lambda t_0$ &  $\bar{\lambda}$ & $M$ & $\chi^2$ \\
 \hline
 \hline
 0.03 & N/A & 0.943 & 0.738 & -19.29 & 1049.75\\ 
 \hline
 0 & 1.000 & $0.828 \pm 0.017$ & $0.715\pm 0.012$ & -19.31 & 1035.68 \\
 \hline
 -0.1 & 0.924 & $0.869 \pm 0.016$ & $0.595 \pm 0.009$ & -19.31 &  1034.25\\
 \hline
 -0.5 & 0.841 & $0.914 \pm 0.015$ & $0.490 \pm 0.006$ & -19.32 & 1032.91 \\
 \hline
 -1 & 0.802 & $0.937 \pm 0.014$ & $0.496\pm 0.005$ & -19.32 & 1032.64\\
 \hline
 -1.2 & 0.793 & $0.943 \pm 0.014$ & $0.507\pm 0.005$ & -19.32 & 1032.63\\
 \hline
 -2 & 0.770 & $0.959\pm 0.013$ & $0.558 \pm 0.005$ & -19.32 & 1032.72\\
 \hline
 -10 & 0.730 & $0.987\pm 0.013$ & $0.844 \pm 0.006$ & -19.33 & 1033.40\\
 \hline
 -$\infty$ & 0.717 & $0.998\pm 0.013$ & $1.177\pm 0.008$ & -19.33 & 1033.85\\
\hline
\end{tabular}
\caption{The scale factor modifier $\beta_\epsilon$ and the best values for $H_\lambda t_0$, the scaled parameter $\bar\lambda$, the absolute magnitude of the supernovae $M$, and the resulting best-fit value of $\chi^2$ for eight nonzero representative values of $\epsilon$. Errors are included for $H_\lambda t_0$ and $\bar\lambda$, except in the case $\epsilon=0.03$, where the errors would be asymmetric, but the $\chi^2$ value is already awful. Note that all values $\epsilon<0$ give a slightly better fit than $\epsilon = 0$, the standard $\Lambda$CDM value. The numerical minimum of $\chi^2$ is at $\epsilon=-1.2$. \label{tab:DataTable}}
\end{center}
\end{table}

\section{Conclusion}\label{Sec:Conclusion}

Under the assumption of a flat Universe, we have studied the scale factor of the Universe in $f(R,T)$ gravity where $f(R,T)=R+\lambda T^{\epsilon}$. Similarly to \cite{Siggia_2025}, for models when $\epsilon\leq0$ we found that the matter-dominated era transitions to exponential growth for the present-day cosmology when considering the analysis of \cite{Fisher_2019}. We have also compared our $f(R,T)$ model predictions for luminosity distance versus redshift with SNe Ia data \cite{Scolnic_2018}. Our numerical results in Table \ref{tab:DataTable} show that all models with $\epsilon<0$ fit the data slightly better than the standard $\Lambda$CDM  model. The best fit for the data occurs near $\epsilon=-1.2$. For $\epsilon>0$ it was shown that the models diverge from $\Lambda$CDM and quickly become a terrible fit for an accelerating Universe.

From our analysis, it is clear that  for $\epsilon \le 0$, our model fits the data as well as $\Lambda$CDM. However, we currently cannot discern which model causes the acceleration. As shown in Fig.~\ref{fig:Apparent_Magitnude}, the deviation from standard $\Lambda$CDM is most noticeable at high redshift. If only SNe Ia data are considered, the collection of more data at high $z$ is necessary to differentiate between these models.

In future work, we plan to examine how our model fits BAO and CMB data. We suspect that this will be a more discerning discriminator between models. 
\\
\section*{Data Availability}

The data supporting the findings of this article are openly available \cite{Scolnic_2018}. The complete data set and the binned data can be found at doi:{\href{https://archive.stsci.edu/doi/resolve/resolve.html?doi=10.17909/T95Q4X}{10.17909/T95Q4X}}\;\cite{Scolnic_doi}. The code used to analyze these data for our models is openly available at {\href{https://github.com/VRSiggia/Analysis_SNe_Ia_Data}{https://github.com/VRSiggia/Analysis\_SNe\_Ia\_Data}.

\medskip

\end{document}